\documentclass[11pt,a4wide]{article}

\usepackage{amsfonts}
\usepackage{amsbsy} 
\usepackage{epsfig}
\usepackage{latexsym}
\input amssym.def
\input amssym.tex
\def\a{{\mathcal{M}}}
\advance \topmargin by -\headheight
\advance \topmargin by -\headsep
\evensidemargin \oddsidemargin
\advance\hoffset by -3mm  
\advance\voffset by  8mm  
\def\bbbc{{\mathchoice {\setbox0=\hbox{$\displaystyle\rm C$}\hbox{\hbox
to0pt{\kern0.4\wd0\vrule height0.9\ht0\hss}\box0}}
{\setbox0=\hbox{$\textstyle\rm C$}\hbox{\hbox
to0pt{\kern0.4\wd0\vrule height0.9\ht0\hss}\box0}}
{\setbox0=\hbox{$\scriptstyle\rm C$}\hbox{\hbox
to0pt{\kern0.4\wd0\vrule height0.9\ht0\hss}\box0}}
{\setbox0=\hbox{$\scriptscriptstyle\rm C$}\hbox{\hbox
to0pt{\kern0.4\wd0\vrule height0.9\ht0\hss}\box0}}}}

\makeatletter
\@addtoreset{equation}{section}
\makeatother

\begin{document}
\hfuzz=100pt
\title{{\Large \bf{Classical Boundary-value Problem in Riemannian
Quantum Gravity and Self-dual Taub-NUT-(anti)de
Sitter Geometries}}}
\author{\\M M Akbar\footnote{Email:M.M.Akbar@damtp.cam.ac.uk}\ \,\,\& P
D D'Eath\footnote{E-mail:P.D.DEath@damtp.cam.ac.uk}  
\\
\\
Department of Applied Mathematics and Theoretical Physics,
\\ Centre for Mathematical Sciences,
\\ University of Cambridge, 
\\ Wilberforce Road,
\\ Cambridge CB3 0WA,
\\U.K.}
\date{\today}
\maketitle
\begin{center}
DAMTP-2002-26
\end{center}
\begin{abstract}
The classical boundary-value problem of the Einstein field equations is studied with an arbitrary cosmological constant, in the case of a compact ($S^{3}$) boundary given 
a biaxial Bianchi-IX positive-definite three-metric, specified by two 
radii $(a,b).$  For the simplest, four-ball, topology of the manifold
with this boundary, the regular classical solutions are found within
the family of
Taub-NUT-(anti)de Sitter metrics with self-dual Weyl
curvature. For arbitrary choice of positive radii $(a,b),$ we find
that there are three solutions for the infilling geometry of this type.
We 
obtain exact
solutions for them and for their Euclidean actions. 
The case of
negative cosmological constant is
investigated further. For reasonable
squashing of the three-sphere, all three infilling solutions have real-valued
actions which possess a ``cusp catastrophe'' structure with a non-self-intersecting 
``catastrophe manifold'' implying that the dominant contribution comes
from the unique real positive-definite solution on the ball. The
positive-definite solution
exists even for larger deformations of the three-sphere, as long as
a certain inequality between
$a$ and $b$ holds. The action of this solution is proportional to $-a^{3}$ for large $a\, (\sim b)$ and hence larger radii are favoured.
The same boundary-value problem with more complicated interior topology containing a
``bolt'' is investigated in a forthcoming paper.
\end{abstract}
\noindent
\section{Introduction}
In quantum gravity, as treated by the combined approaches of the Dirac 
canonical quantization and its dual, the Feynman path integral \cite{fm},
one studies (for example) the amplitude for $n$ disconnected 
compact three-surfaces to have given spatial three-metrics, 
respectively $h^1_{ij}, h^2_{ij}, ...,h^n_{ij}.$  The amplitude 
is given formally by a path integral.  In the simplest case 
$n = 1$, with only one connected three-surface, which could then be
regarded as a spatial cross-section of a cosmological model, the
amplitude is given by the `no-boundary' or `Hartle-Hawking' state 
\cite{HH}
\begin{equation} 
{\Psi}_{HH}(h_{ij})
={\int}{\cal D}g_{{\mu}{\nu}}{\exp}(-I_{E}[g_{{\mu}{\nu}}]), \label{1}
\end{equation}   
where the integral is over Riemannian four-geometries $g_{{\mu}{\nu}}$
on compact manifolds-with-boundary, where the three-metric induced on
the boundary agrees with the prescribed $h_{ij}$ above. Here $I_{E}$
denotes the Euclidean action \cite{Gibbons:1976ue} of the four-dimensional
configuration, as in Eq.(\ref{29}) below.

Naturally, in considering semi-classical approximations to this
integral, one is led first to study the classical Riemannian
boundary-value problem: namely, to find one (or more) Riemannian
solutions to the classical Einstein field equations, possibly
including a cosmological constant ${\Lambda},$ obeying
\begin{equation}
R_{{\mu}{\nu}}={\Lambda}g_{{\mu}{\nu}}, \label{2}
\end{equation}
which are smooth on the interior manifold, and which agree with the
spatial three-metric $h_{ij}$ at the boundary. Below, we shall see
examples in which (different) such classical solutions exist, for a
class of choices of the three-metric $h_{ij}$ on a boundary diffeomorphic 
to the three-sphere $S^{3}$, on a manifold-with-boundary with the
simplest possibility of a four-ball topology. 

Here, the boundary-value problem is studied within the
class of biaxial Bianchi-IX models \cite{EGH,GP1}, which
may be written locally in the form
\begin{equation}
ds^2 = dr^2 + a^2(r)({\sigma}_1^{~2}+{\sigma}_2^{~2})
              + b^2(r){\sigma}_3^{~2}. \label{3}
\end{equation}
Here ${\sigma}_i,{\;}i=1,2,3,$ denote the left-invariant one-forms on
$S^{3}$ (see, for example, \cite{GP1}). The boundary 3-metric at a given value $r_{0}$ of
$r$ is then a Berger three-sphere \cite{ped,sak}, with intrinsic three-metric
\begin{equation}
ds^2 = a^2(r_{0})({\sigma}_1^{~2}+{\sigma}_2^{~2})
              + b^2(r_{0}){\sigma}_3^{~2}. \label{4} 
\end{equation}
Subject to condition (\ref{2}), the metrics (\ref{3}) are known in a closed form --
they are the well-known family of Riemannian Taub-NUT-(anti)de Sitter metrics
\cite{Carter,EGH,GP1}. For the topologically simplest case that we are
studying in this paper, we insist that the $S^{3}$ with the given
intrinsic metric (\ref{4}) bound a four-ball with 
smooth four-metric; this corresponds to requiring the Taub-NUT-(anti)de
Sitter metrics to close with a regular
``nut'' and to have half-flat Weyl curvature. We find that the
problem can be translated into an algebraic system of degree three which
can be solved exactly to find the possible infilling geometries of
this type and their actions for \emph{any} boundary data $a,b$. 
Depending on
$a,b,$ one may therefore have three real roots, or
one real root together with one complex conjugate pair, for this third-degree
equation. A similar study was carried out in
\cite{Louko}, for large three-volume and small anisotropy, assuming a
positive cosmological constant. However, we
have been able to find all complex- and real-valued self-dual Taub-NUT-(anti)de
Sitter solutions on the 4-ball and their actions for arbitrary
values of $(a,b)$ for both positive and negative cosmological constants. The
case of zero cosmological constant can be obtained by
taking the cosmological constant to zero (equivalently) from either the positive 
or the negative direction. 

Self-dual Riemannian Einstein spaces with a negative cosmological
constant and of
biaxial Bianchi-IX type have also been studied, for example, by
Pedersen \cite{ped}, in connection with the conformal boundary-value
problem with a three-sphere (topologically) at infinity. The more
general case of generic (non-biaxial) Bianchi-IX models has been
treated in this context by Hitchin \cite{Hit} and by Tod \cite{Tod}.

Since a strictly
negative cosmological constant ${\Lambda}$ is needed to have any hope
of incorporating gauge theories of matter with local supersymmetry in
a four-dimensional field theory \cite{PD,WB}, we have
given a more detailed analysis for $\Lambda<0$; the real and complex
solutions have been classified completely in terms of the boundary
values $a,b$,
and the numerical behaviour of the solutions and Euclidean action $I_{E}$
has been
investigated in greater detail.
\section{Einstein Biaxial Bianchi-IX Metrics of Riemannian Signature}
The general form of a biaxial Bianchi-IX four-metric is given by 
Eq.(\ref{3}). Such metrics are invariant under the group
action of $SU(2) \times U(1)$, whose Lie algebra is isomorphic to
that of $U(2)$. When one further
imposes the Einstein equations with a $\Lambda$ term, one arrives at
the two-parameter Taub-NUT-(anti)de Sitter family \cite{Carter,EGH,GP1} :
\begin{equation}
ds^2=\frac{\rho^{2} - L^{2}}{\Delta} d\rho^{2}+ \frac{4
L^{2}\Delta}{\rho^{2}-L^{2}}(d\psi+\cos \theta\,
d\phi)^{2}+(\rho^{2}-L^{2})(d\theta^{2}+\sin^{2}\theta \,d\phi^{2}), \label{Taub}\label{5}
\end{equation}
where
\begin{equation}
\Delta=\rho^{2}-2M\rho +L^{2}+\Lambda \left( L^{4}+2
L^{2}\rho^{2}-\frac{1}{3}\rho^{4}\right).
\end{equation}
Here $L$ and $M$ are the two parameters and $0\leq\theta\leq\pi$,
$0\leq\phi\leq2\pi$, $0\leq\psi\leq4\pi/k$ ($k$ is a natural
number). When $k=1$, the surfaces of constant $\rho$ are topologically
$S^{3}$.
The general form (\ref{5}), however, is only valid for a coordinate patch for
which $\Delta \ne
0$. In general $\Delta$ will have four roots. At the roots the metric
degenerates to that of a round $S^{2}$, and each such root therefore corresponds to
a two-dimensional set of fixed points of the Killing vector field
$\partial/{\partial \psi}$. However, if a root occurs at $\rho=|L|$, the
fixed-point set is zero-dimensional (as the two-sphere then
collapses to a point). Such two- and zero-dimensional fixed
point sets have been given the names ``bolts'' and ``nuts''
respectively \cite{GH}. The coordinate $\rho$
ranges continuously from a root of $\Delta$ until it
encounters another root of 
$\Delta$, if there is any; otherwise $\rho$ ranges from the root to
infinity. In general, the
bolts of the above Taub-NUT family of metrics are not regular
points of the metric. For them to be regular,
the metric has to close smoothly at the bolts, for which the
condition is \cite{Page}:
\begin{equation}
\frac{d}{d\rho}\left(\frac{{\Delta}}{\rho^{2}-L^{2}}
\right)_{(\rho=\rho_{bolt})}=\frac{1}{2kL}.\label{7}
\end{equation}
For finite $L\ne 0$ and $k=1$, condition (\ref{7}) leads to the
self-dual Taub-NUT-(anti)de Sitter metric (see
below) and the Taub-Bolt-(anti)de Sitter metric \cite{EGH,GP2}, which
 reduce to the
Euclidean Taub-NUT \cite {Hawk} and 
the Taub-Bolt metrics \cite{Page} respectively for $\Lambda=0$. For the limiting cases
 $L \rightarrow \infty$ and $L=0$, for which (\ref{5}) is not
 well-defined, one
 can obtain
 regular solutions by suitable coordinate transformations and assigning correct
 periodicities to the coordinate parametrizing the $S^1$ fibre. For
example, for $\Lambda=0$, the Eguchi-Hanson metric \cite{EH} and the
 Schwarzschild metric can be obtained from (\ref{5}) in
these limits \cite{Page}. In the next section we will
 encounter more examples of Bianchi-IX metrics that arise at these
 limits as we discuss the self-dual
 Taub-NUT-(anti)de Sitter solutions. For a recent
discussion on obtaining different biaxial Bianchi-IX metrics as
various limits of (\ref{5}) see \cite{CGP}.
\subsection{Self-dual Weyl tensor and Regularity}
It will become evident below that the regular, positive-definite geometries of this type
on a four-ball interior are Weyl half-flat, that is, having either
self-dual or anti-self-dual Weyl curvature \cite{AHS,GP2} which requires:
\begin{equation}
M=\pm L(1+\frac{4}{3}\Lambda L^{2}) \label{8}
\end{equation}
For either sign, such metrics are often called (by physicists)
half-flat -- a name which we will be using in this paper.

It is easy to check that, when the metric has a nut -- or equivalently $\Delta$ has
a (double) root at $\rho= L$ (or $\rho= -L$) -- then the two parameters $L$ and $M$ are related by
\begin{equation}
M= \pm L(1+\frac{4}{3}\Lambda L^{2}),\label{9}
\end{equation}
which is precisely the condition above for self-duality of the Weyl
tensor. One can further check that Eq.(\ref{7}) is automatically satisfied
at the nut of the metric (\ref{5}). Therefore, within the Taub-NUT-(anti)de Sitter family, self-duality of the Weyl tensor will imply
regularity at the origin.

Without loss of generality, we will
work with the positive sign of Eq.(\ref{8}) (and (\ref{9})). The condition (\ref{8}) of
half-flatness (or condition (\ref{9}) of nut-regularity) implies that
$\Delta$ simplifies:
\begin{equation}
\Delta=(\rho - L)^{2}-\frac{1}{3}\Lambda(\rho+3 L)(\rho-L)^{3}.
\end{equation}
 One can note that the terms involving $\Lambda$
vanish as $\rho \rightarrow L$, and that the metric near the origin is
that of the $\Lambda=0$ case, namely Hawking's
(Taub-NUT) solution \cite{Hawk}:
\begin{equation}
ds^2=\left(\frac{\rho + L}{\rho - L}\right) d\rho^{2}+4 L^{2}\left(\frac{\rho - L}{\rho + L}\right) (d\psi+\cos \theta
d\phi)^{2}+(\rho^{2}-L^{2})(d\theta^{2}+\sin\theta^{2} d\phi^{2})\label{11}
\end{equation}
Apart from the double root at $\rho=L$, $\Delta$ has two more roots or ``bolts'' at $\rho=\pm
\sqrt{(4 L^{2}+3 /\Lambda)}-L$. Beyond these two roots $\Delta$ is
negative. So, the permissible range of $\rho$ is from $L$ to
$\sqrt{(4 L^{2}+3 /\Lambda)}-L$ (when $L$ is positive) or from $L$ to $-\sqrt{(4
L^{2}+3 /\Lambda)}-L$ (when $L$ is negative). For a complete nonsingular
metric, we have to check whether such points are regular, i.e., whether
\begin{equation}
\left(\frac{2}{3}\frac{(L-\rho)(\rho^{2}+3L\rho+4 L^{2})+3 L}{(\rho+L)^{2}}\right)_{(\rho=\rho_{bolt})}=\frac{1}{2L}
\end{equation}
which is identically satisfied at the nut (at
$\rho=L$) as we
have seen already. However,
for $\rho=
\sqrt{(4 L^{2}+3/ \Lambda)}-L$, which is positive definite and greater
than $L$, this would require:
\begin{equation}
\frac{4}{3} \Lambda L- \frac{2}{3} \sqrt{(4\Lambda^{2} L^{2}+3
\Lambda)}=\frac{1}{2L}
\end{equation}
which can only happen for $L$ negative
($=-\frac{1}{8}\sqrt{6/\Lambda}$) and hence we find a
contradiction. 
(In other words, this would 
give a
regular bolt at
$\rho=\frac{7}{8}\sqrt{6/\Lambda}$ -- on the
``other side'' of $\rho=L$ which is negative, and hence cannot be reached continuously
starting from $L$.) 
The same argument applies
for the other root at $\rho=
-\sqrt{(4 L^{2}+3/ \Lambda)}-L$.
Therefore, for positive cosmological constant, the half-flat
(equivalently, regular-nut) solution, has a singular bolt at finite
$\rho$ which exists and cannot be made
regular for any finite value of the parameter $L\ne 0$.
However, in the limit
$L \rightarrow 0$, one can obtain the standard round metric on $S^{4}$
by suitable coordinate transformations; this is regular everywhere and
has a regular nut at each of
the two poles. 
For the $L \rightarrow
\infty$ limit, on the other hand, one can 
obtain the regular Fubini-Study metric on $\bbbc P^{2}$. 
This is regular everywhere and has a regular nut at the origin and a
regular bolt at infinity \cite{GP1}. 

The case with a negative cosmological constant is different.
Writing $\Lambda= - \lambda $ ($\lambda $ is positive), one has
\begin{equation}
\Delta=(\rho - L)^{2}+\frac{1}{3}\lambda(\rho+3 L)(\rho-L)^{3}.
\end{equation}
$\Delta$ now has two roots at at $\rho=L$ and two others at $\rho=\pm
\sqrt{(4 L^{2}-3 \lambda)}-L$, which are beyond the permissible range of
$\rho \ge L$ ($L$ positive) and $\rho \le L$ ($L$ negative). So, for a fixed negative $\Lambda$, the one-parameter family of half-flat solutions of
the Taub-NUT-anti de Sitter type are necessarily regular for
$\rho \in [L,\infty)$ for any finite, non-zero value of $L$. 
Using similar coordinate transformations as in the case of positive cosmological constant, one can
obtain the canonical metric on $H^4$ and the Bergman metric on
$\overline{\bbbc P^{2}}$ in the limits of $L \rightarrow 0$ and $L
\rightarrow \infty$ respectively, both of which are nut-regular at the
origin.

So, for both $\Lambda >0$ and $\Lambda < 0$, in this half-flat
Taub-NUT-(anti) de Sitter family any hypersurface of constant $\rho$
(which is a Berger sphere with a given 3-metric) bounds a four-ball
with a regular 4-metric. The singular bolt at  $\rho=
\pm \sqrt{(4 L^{2}+3/ \Lambda)}-L$, in the case of $\Lambda>0$,
poses no problem as the Berger sphere lies between the
(regular) nut and the (singular) bolt when we are interested in
filling the Berger sphere with positive-definite regular metrics. It remains to see which classical 4-metrics of Taub-NUT-(anti)de
Sitter type can be given on the
four-ball inside a given Berger sphere. In other
words, how many members in
this one-parameter ($L$) family of metrics are there for which a given
Berger sphere
is a possible hypersurface at some constant $\rho$? 

In addressing this
question we can
include complex
solutions in the interior of the 3-sphere. For this we let $\rho$
and $L$ and hence the 4-metric be in
general complex-valued as long as the 4-metric induces the
prescribed positive-definite 3-metric on the boundary and is
non-singular in the interior.\footnote{For a detailed discussion
on such complex-valued metrics on real manifolds see \cite{Louko}
and references therein.} We will return to this issue in the next section. Real positive-definite infilling
4-metrics therefore form a subclass of the complex-valued
solutions to this boundary-value problem and do not necessarily exist for
arbitrary boundary data. However, it is possible to find the necessary
and sufficient condition on the boundary data for the existence of
real positive-definite solutions as we will see in Sections 4 and 5.
\section{Infilling Geometries and their Action}
Following the discussion above, the problem of finding classical infilling regular/(anti)self-dual
Einstein metrics
of Taub-NUT-(anti)de Sitter type on the four-ball bounded by a typical
Berger sphere with two radii $(a,b)$ can be translated into solving the
following two
equations in $\rho$ and $L$: 
\begin{equation}
a^{2}-\rho^{2}+L^{2}=0, \label{15}
\end{equation}
and
\begin{equation}
b^{2}(\rho^{2}-L^{2})-4\,{L}^{2}\left (\left (\rho-L\right )^{2}-\frac{1}{3}\,\Lambda\,\left (\rho+3
\,L\right )\left (\rho-L\right )^{3}\right )=0. \label{16}
\end{equation}
These are polynomial equations of
degree two and six in the variables $\rho$ and $L$ and hence, by
B\'{e}zout's Theorem (see, for example, \cite{Reid}), would intersect at at most twelve points in $\bbbc^{2}$. However, trying to solve Eq.(\ref{15})-(\ref{16}) explicitly for
$\rho$ and $L$ is not
very easy and not to be recommended. However, there is a more
``symmetric'' approach, which will enable us to
find explicit solutions, and first of all enables us to count the
number of infilling geometries.
\subsection{Number of Infilling Geometries}
{\bf{Theorem:}} \emph{For any boundary data $(a,b)$, where $a$ and
$b$ are the radii of two equal and one unequal axes of a Berger
3-sphere (squashed
3-sphere with two axes equal) respectively, there are precisely three
(modulo ``orientation'') self-dual
Taub-NUT-(anti)de Sitter geometries which
contain the 3-sphere as their boundary.\footnote{As we are dealing
with finite $L \ne 0$ for which (\ref{5}) is
well-defined, self-dual metrics of biaxial Bianchi-IX type (admitting
$S^3$-foliations) arising as 
limiting cases ($L \rightarrow 0$ and $L \rightarrow \infty$) of the self-dual Taub-NUT-(anti)de
Sitter -- as discussed in
Section $2$ -- are
precluded from the considerations here. Depending on two radii of the
Berger sphere, possible infilling solutions
of such metrics can easily be included when
one considers the larger class of (self-dual) Bianchi-IX type
metrics on the four-ball.}}
\\
{\bf{Proof:}} Note that the system of equations (\ref{15})-(\ref{16}) admits the
discrete symmetry $(\rho, L)
\leftrightarrow (-\rho, -L)$.
(This is what we mean by ``orientation'').
Make the substitution:
\begin{equation}
\begin{array}{rcl}
x&=&\rho+L\\
y&=&\rho-L.
\end{array} \label{17}
\end{equation}
The problem now reduces to solving
\begin{equation}
a^{2}=xy \label{18}
\end{equation}
and
\begin{equation}
b^{2}=\frac{y}{3x}\left (x-y\right )^{2}\left (3-2\,\Lambda\,xy+\Lambda\,{
y}^{2}\right ).\label{19}
\end{equation}
Note that this preserves the discrete symmetry, $(\rho, L)
\leftrightarrow (-\rho, -L)$ in $(x,y) \leftrightarrow
(-x,-y)$.
Substitution of $x$ in Eq.(\ref{19}) now gives the univariate equation:
\begin{equation}
\Lambda y^{6}+(3-4 a^{2} \Lambda) y^4 + (5 a^{4} \Lambda -6 a^{2}) y^2
-(3a^{2}b^{2}-3a^{4}+ 2\,a^{6}\Lambda)=0 \label{20}
\end{equation}
which is a cubic equation in $y^{2}$. The six solutions for $y$
therefore will appear in pairs of opposite signs. Since
$a^{2}$ is positive, this implies that the six corresponding
solutions of $(x,y)$ are of the form $(x_{1},y_{1})$, $(x_{2},y_{2})$, $(x_{3},y_{3})$
and $(-x_{1},-y_{1})$, $(-x_{2},-y_{2})$, $(-x_{3},-y_{3})$. 

Since the set of solutions have the symmetry  $(x_{i},y_i{}) \rightarrow
(-x_{i},-y_{i})$, by applying the transformation $(x,y) \rightarrow
(-x,-y)$, we would obtain no new solutions
for $(x,y)$ (hence for $(\rho,L)$) and would reproduce the same set. Therefore there are
six points in $\bbbc^{2}$ where the two
polynomials meet, i.e., six solutions for $(\rho, L)$ which are related by
the reflection symmetry $(\rho, L)
\leftrightarrow (-\rho, -L)$. Hence the number of geometries
modulo orientation for any
given boundary data $(a,\,b)$ is three. \,\,$\Box$
\subsection{Real/Complex Roots and Real Infilling Metrics}
It is more convenient to rewrite Eq.(\ref{20}) for $z=y^{2}$ in the form 
\begin{equation}
\Lambda z^{3}+(3-4 A \Lambda) z^2 + (5 A^{2} \Lambda -6 A) z
-(3AB-3A^{2}+ 2\,A^{3}\Lambda)=0 \label{21}
\end{equation}
where we have denoted $(a^{2},b^{2})$ by $(A,B)$.  
It is easy to see that the real solutions for $(x,y)$ (and hence for
$(\rho,L)$) come from the {\it{positive}} roots of
Eq.(\ref{21}). Depending on the boundary data $(A,B)$, the number of
positive solutions for $z$ can range from zero to
three. However, a positive solution of $z$ and, hence a real solution $(\tilde{\rho},\tilde{L})$, would not
necessarily correspond to a {\it{real positive-definite infilling geometry}}. For it to qualify
as a regular real solution in the interior, the metric should be
well-defined for $\rho$ taking values on the real line from $\tilde{L}$
to $\tilde{\rho}$. As we discussed in
Section $2.1$, this can happen if and only if $\tilde{\rho}$
and $\tilde{L}$ have the same sign and $|\tilde{\rho}|>|\tilde{L}|$. The
latter condition is automatically satisfied as a consequence of $a^2$
being positive
irrespective of whether the signs of $\tilde{\rho}$
and $\tilde{L}$ are similar or not. 
If
$\tilde{\rho}$ and $\tilde{L}$ have opposite signs, the metric will
not be positive definite for values of $\rho$ within the interval
$(\tilde{L},-\tilde{L})$ on the real line, and
will be singular at $\rho=-\tilde{L}$. Therefore, it would not qualify
as a real positive-definite solution. Such solutions for $(\tilde{\rho},\tilde{L})$
should be considered to correspond to complex metrics where $\rho$ is generally complex-valued in the
interior and
real-valued on the given $S^{3}$ boundary and at $\tilde{L}$. (One can choose 
contours for $\rho$ in the complex plane which avoid the
singularity on the real line at $-\tilde{L}$ and ensure regularity at
$\tilde{L}$; see discussions in \cite{Louko}.) We will discuss the dependence of positivity (and complexity) of
$z$ on the boundary data $(A,B)$ in detail for $\Lambda<0$ in Section
$5$ to explore the structure of (real) solutions and their actions
and shall return to this issue. We will show that for a given
boundary data $(A,B)$ the real regular
infilling solution is unique and exists when a certain inequality holds
between $A$ and $B$.

For negative or complex roots of Eq.(\ref{21}), the interior metrics are necessarily complex-valued.
\subsection{Explicit Solutions}
We now write down explicit solutions for our boundary-value
problem. Rewrite Eq.(\ref{21}) with the conformal rescaling
\begin{equation}
\begin{array}{rcl}
A |\Lambda| &\rightarrow& A,\\
B |\Lambda| &\rightarrow& B,\\
z |\Lambda| &\rightarrow& z,\\
\end{array}
\end{equation}
so that $A,B$ are now dimensionless. For positive cosmological constant, this gives
\begin{equation}
f(z):= z^{3}+(3-4A) z^2 - (5 A^{2} -6 A) z +(3AB-3A^{2}+2 A^{3})=0,\label{23}
\end{equation}
and, for negative cosmological constant,
\begin{equation}
g(z):= z^{3}-(4A +3) z^2 + (5 A^{2} +6 A) z +(3AB-3A^{2}-2 A^{3})=0\label{24}.
\end{equation}
Since we will be more interested in the case of negative cosmological constant,
we only write down the solutions of $g(z)=0$. These are
\begin{equation}
z_{1}=\frac{1}{6}\,\sqrt [3]{Q}+{\frac {\frac{2}{3}\,{A}^{2}+4\,A+1}{\sqrt [3]{Q}}}+1+
\frac{4}{3}\,A,\label{sol1}
\end{equation}
\begin{equation}
z_{2}=-\frac{\sqrt [3]{Q}}{12}-{\frac {\frac{1}{3}\,{A}^{2}+2A+3}{\sqrt [3]{Q}}}+
1+\frac{4}{3}\,A+\frac{i}{2}\,\sqrt {3}\left (\frac{1}{6}\,\sqrt [3]{Q}-{\frac {\frac{2}{3}\,{A}
^{2}+6\,A+6}{\sqrt [3]{Q}}}\right ),\label{sol2}
\end{equation}
and
\begin{equation}
z_{3}=z_{2}^{*},\label{sol3}
\end{equation}
where 
\begin{equation}
Q=72\,{A}^{2}+8\,{A}^{3}+216\,A-324\,AB+216+36\,\sqrt {AB\left (81\,AB-108-108
\,A-36\,{A}^{2}-4\,{A}^{3}\right )}
\end{equation}\\
The solutions of $f(z)=0$ can be
found by just flipping the signs of $A$ and $B$.
It is important to note here that, the appearance of explicit factors
of $i$ in the expressions (\ref{sol1})-(\ref{sol3}) for the roots may not be an
accurate guide as to whether a root is real or complex. In Section $4$, we will see how to 
analyse
the roots of $g(z)$ (in a way that would enable one to
determine whether the roots are complex or real, for different values of $(A,
B)$) without invoking the explicit solutions (\ref{sol1})-(\ref{sol3}). However,
we do need the explicit solutions in order to calculate the action for such
geometries.
\subsection{Action Calculation}
The Euclidean action is given by \cite{Gibbons:1976ue}:
\begin{equation}
8 \pi \,G \,I_{E} = -\int_{\a} d^4 x \sqrt{g} \,\left(\frac{1}{2} R - 
\Lambda\right)- \int_{\partial \a} d^3x \sqrt{h}\, K \label{29}
\end{equation}
where $\a$ is the Riemannian 4-manifold with boundary $\partial
\a$, and the 4-geometry $g_{\mu\nu}$ agrees with the specified three-metric
$h_{ij}$ on the boundary, i.e., $g_{{ij}|_{\partial \a}}=h_{ij}$. $K$ is
the trace of the extrinsic curvature $K_{ij}$ of the boundary.

For a classical solution, obeying the Euler-Lagrange equations of
the action (\ref{29}) (the Einstein equations $R_{\mu\nu}= \Lambda
g_{\mu\nu}$), one has
\begin{equation}
8 \pi \,G \,I_{E} = -{\Lambda}\int_{\a} d^4 x \sqrt{g} - \int_{\partial \a} d^3x \sqrt{h}\, K.
\end{equation}
Now $\int_{\a} d^4 x \sqrt{g}$ is just the volume of $\a$, which for
the self-dual Taub-NUT-(anti)de Sitter metric gives
\begin{equation}
-{\Lambda}\int_{\a} d^4 x \sqrt{g}=-\frac{32{\Lambda}}{3}\pi^2\,L\,(\rho-L)^2\,(\rho+2L)=-\frac{8{\Lambda}}{3}\,{\pi}^{2}\left
({y}^{2}-A\right )\left ({y}^{2}-3\,A\right ). \label{31}
\end{equation}
The trace of the extrinsic curvature is
$(2\frac{a'}{a}+\frac{b'}{b})/N$ for any metric of the form
\begin{equation}
ds^2 = N(r)^{2}dr^2 + a^2(r)({\sigma}_1^{~2}+{\sigma}_2^{~2})
              + b^2(r){\sigma}_3^{~2}
\end{equation}
giving
\begin{equation}
 - \int_{\partial \a} d^3x \sqrt{h}\,
   K= -\frac{1}{N}\left(2\frac{a'}{a}+\frac{b'}{b}\right)(16 \pi^{2} a^{2}b).
\end{equation}
For the self-dual Taub-NUT-(anti)de Sitter metric the surface term in Eq.(\ref{29}) is therefore 
\begin{equation}
 - \int_{\partial \a} d^3x \sqrt{h}\,
   K=\frac{8}{3}{\pi}^{2}\,{\frac {\left ({y}^{2}-A\right )\left (\Lambda\,{y}^{4}+\Lambda\,A{y}^{2}+3\,{y
}^{2}+9\,A-8\,\Lambda\,{A}^{2}\right )}{A}}.
\end{equation}
The total action is then:
\begin{equation}
8 \pi \,G \,I_{E}=\frac{8}{3}\,{\frac {{\pi}^{2}\left ({y}^{2}-A\right )\left (-5\,\Lambda\,{A}
^{2}+9\,A+3\,{y}^{2}+\Lambda\,{y}^{4}\right )}{A}}
\end{equation}
Using Eq.(\ref{21}), $I_{E}$ can be further simplified to give:
\begin{equation}
8 \pi \,G \,I_{E}=\frac{8}{3}\,{\pi}^{2}\left (7\,\Lambda\,{A}^{2}-12\,A+12\,z+3\,\Lambda\,{z}^{
2}-10\,\Lambda\,Az+3\,B\right ) \label{36}
\end{equation}
which in terms of the scaled variables gives, in the case $\Lambda>0$
\begin{equation}
8 \pi \,G \, (\Lambda I_{E})=\frac{8}{3}\,{\pi}^{2}\left (7\,{A}^{2}-12\,A+12\,z+3\,{z}^{
2}-10\,Az+3\,B\right ) \label{37}
\end{equation}
In the case of a negative cosmological constant, this is just:
\begin{equation}
8 \pi \,G \, (\lambda I_{E})=-\frac{8}{3}\,{\pi}^{2}\left (7\,{A}^{2}+12\,A-12\,z+3\,{z}^{
2}-10\,Az-3\,B\right ) \label{38}
\end{equation}
Clearly, the substitutions (\ref{17}) have led to considerable
simplifications. The action is a quadratic function of $z$ -- the
solution of the third-degree equation -- and hence would, in general, be complex (real)
for $z$ complex (real). Corresponding to every solution
of Eq.(\ref{23}) or Eq.(\ref{24}), one
can calculate the action merely by substitution in Eq.(\ref{37}) or
Eq.(\ref{38}) -- one does \emph {not} need to
find $y$ (or $x$) at all to evaluate 
the action. As we discussed in Section $3.2$, not all positive solutions for $z$ correspond to
real positive-definite solutions in the interior. Therefore all real- and complex-valued solutions in the interior
corresponding to the positive solutions for $z$ have real-valued
Euclidean actions. This is because the two limits (here $\tilde{\rho}$ and
$\tilde{L}$) of the integral (\ref{31})
of $\rho$ lie on the real-line. Interestingly, for negative
solutions for $z$, which correspond to $\tilde{\rho}$ and
$\tilde{L}$ being purely imaginary, the actions are real as well. This
is not surprising given that the middle expression of
(\ref{31}) is a quartic polynomial in $\rho$ and $L$.
Only the
complex solutions for $z$ would in general have complex actions.
\section{Vacuum Case}
Before embarking on solutions with non-zero
 cosmological constant, it is desirable
 to understand the $\Lambda=0$ case. This will also
 provide us with an
 example and
 illustrate the utility of the method developed so far.
 For this case, the regular Taub-NUT metrics are given by the Hawking
solution (\ref{11}). The corresponding (quadratic) polynomial
equation and actions are obtained by simply setting $\Lambda=0$ in
Eq.(\ref{21}) and Eq.(\ref{36}) respectively. The two solutions and their actions are
\begin{equation}
z=A\pm\sqrt{AB} \label{solvac}
\end{equation}
and
\begin{equation}
I_{E}= \frac{\pi}{G} (B \pm 4 \sqrt{AB}). \label{actvac}
\end{equation}
Following the discussions in Section $3.2$, it is easy to check that the
negative-sign solution of (\ref{solvac}) corresponds to a real
positive-definite metric on the four-ball provided that 
\begin{equation}
a>b. \label{vac}
\end{equation}
This can be confirmed by solving
for $\rho$ and $L$ directly (which is possible in this case). For the
negative and positive signs of Eq.(\ref{solvac}) they are
\begin{equation}
\rho=\frac{a}{2}\frac{(2a-b)}{\sqrt{a^2-ab}},\,\,\,\,\,L=\frac{b}{2}\sqrt{\frac{a(a-b)}{a-b}},
\end{equation}
and
\begin{equation}
\rho=\frac{a}{2}\frac{(2a+b)}{\sqrt{a^2+ab}},\,\,\,\,\,L=-\frac{b}{2}\sqrt{\frac{a(a+b)}{a+b}}.
\end{equation}
(Two other solutions are obtained by changing orientation.)
Note that,
in contrary to the statement made in \cite{Louko}, positive-definite real
infilling solutions do not exist for arbitrary
boundary data although the actions (\ref{actvac}) are real-valued for any value of $A$ and $B$. The
issue of positive-definite real solutions on the four-ball will be made
clearer in the following sections where we discuss the case of non-zero
cosmological constant (which includes $\Lambda=0$ as a special case).
\section{Solutions for $\Lambda < 0$ and their Actions}
We now treat the case of a negative cosmological constant in greater
detail and systematically discuss the structure of the solutions of 
\begin{equation}
g(z):= z^{3}-(4A +3) z^2 + (5 A^{2} +6 A) z +(3AB-3A^{2}-2 A^{3})=0 \label{44}
\end{equation}
as functions of boundary data $(A,B)$. Recall that
$(\rho,L)$ can have a real solution only when $z$ is positive. Therefore, by studying the solutions of Eq.(\ref{44}), we are
able to see the real and complex solutions of this boundary-value
problem quite readily. However, before doing so we discuss the
possibility of real positive-definite infilling solutions.
\subsection{Real, Positive-definite Infilling Solutions}
For a given boundary data $(A,B)$, let the infilling solutions be
denoted by $(\tilde{\rho},\tilde{L})$. Recall from
our discussion in Section $3.2$ that
a real positive-definite infilling solution exists if $\tilde{\rho}$
and $\tilde{L}$ have like signs (and if $|\tilde{\rho|} >
|\tilde{L}|$ which holds automatically). Assuming without any loss of
generality, that both $\tilde{\rho}$ and $\tilde{L}$
are positive, the equivalent
requirement in the variables $x$, $y$ is $x > y$.
It then follows from Eq.(\ref{18}) that the corresponding requirement
on the possible solutions for $z \equiv y^2$ is the following:
\begin{equation}
z < a^2. \label{neweq1}
\end{equation}
Therefore, only positive roots of
Eq.(\ref{23}) or Eq.(\ref{24}) within the interval $ (0,A)$ give
positive-definite real solutions on the four-ball which are regular
everywhere including the origin, i.e., at $\rho= \tilde{L}$. Note that
the requirement (\ref{neweq1}) is independent of the size or sign of
the cosmological constant.

We now analyse the case of negative cosmological constant
and find the necessary and sufficient conditions on the boundary data
for roots of Eq.(\ref{44}) to satisfy the inequality (\ref{neweq1}).
\\
{\bf{Lemma:}} \emph{There is a unique real, positive-definite self-dual Taub-NUT-anti
de Sitter solution on the four-ball bounded by a  given Berger-sphere of
radii $(a,b)$, if and only if}
\begin{equation}
b^2< \frac{1}{3}a^2\,(2a^2|\Lambda|+3). \label{cond}
\end{equation}
{\bf{Proof:}} Recall that, for any function $F(x)$, if $F(a)$
and $F(b)$ have unlike signs, then an odd number of roots of $F(x)=0$ lie
between $a$ and $b$ (see, for example, \cite{BC}). Depending on the
boundary data there are two distinct possibilities for the sign of the
constant term in Eq.(\ref{44}). For $B<
\frac{1}{3}A(2A+3)$,
$g(0)$ is negative and $g(A)=3AB$ and is strictly positive. Therefore, for $B<
\frac{1}{3}A(2A+3)$, there will be
either one root or three roots in the interval $(0,A)$. 
By direct differentiation, one can show that $g'(z)$ and $g''(z)$ are strictly non-negative within the interval $[0,A]$. It therefore follows from  Fourier's
theorem that the number of roots in the
interval $(0,A)$ is one.\footnote{  Fourier's
theorem: If $F(x)$ be a polynomial of degree $n$ and
$F_{1},F_{2},...,F_{n}$ are its successive derivatives, the number
of real roots $R$ which lie between two real numbers $p$ and $q$
($p<q$) are such that $R\le N-N'$,
where $N$ and $N'$ ($N\ge N'$) respectively denote the number of
changes of sign in the sequence $f_{1},f_{2},...f_{n}$, when $x=p$ and
when $x=q$. Also $((N-N')-R)$ is an even number or zero. See, for
example, \cite{BC}.} If, on the other hand, $B\ge
\frac{1}{3}A(2A+3)$, there will be no root within the interval
$(0,A)$. \,\,$\Box$

Note that (\ref{cond}) gives (\ref{vac}) in the $|\Lambda|=0$ limit. There are certain particular features of the coefficients of
$g(z)$ that simplified our analysis. Firstly, it is only in the constant ($z^{0}$) term
that $B$ appears. Also, and more importantly, except
for the constant term the coefficients of
$g(z)$  are either positive or negative
definite for arbitrary $(A,B)$. (In the case of
a positive $\Lambda$, we are not fortunate enough to have
this particular advantage -- the corresponding analysis is therefore just a step
lengthier, although the methodology developed in this section equally
applies. Possible positive-definite real solutions are unique as in the
case of negative cosmological constant.) 

\subsection{Structure of Roots}
We now discuss how the roots of Eq.(\ref{44}) change as the boundary data is
varied. Since for real (complex) roots the action is
real (complex) this would enable us to study the real (complex)
actions. In fact, as we will see below, regions of physical
interest are covered by real solutions of Eq.(\ref{44}).

As is apparent from the previous discussions, we must look carefully at
the two different cases, for which the constant term is positive or negative.
The
condition determining the reality of the roots of $g(z)$ is following
(see, for example, \cite{AS}):
\begin{equation}
\begin{array}{rcl}
B&<&\frac {4}{81}\,\frac{(A+3)^{3}}{A}\;\;\;\;\; \;\;\;\;\;
{\rm (three\;\;real\;\;roots)}\\
B&=&\frac {4}{81}\,\frac{(A+3)^{3}}{A}\;\;\;\;\; \;\;\;\;\;
{\rm (three\;\;real,\;\;at\;\;least\;\;two\;\; equal)}\\
B&>&\frac {4}{81}\,\frac{(A+3)^{3}}{A}\;\;\;\;\; \;\;\;\;\;
{\rm (one\;\; real\;\;two\;\;complex\;\;roots)}
\end{array} \label{realcon}
\end{equation}
However, for the discussion below one further needs to know
Descartes' Rule \cite {BC}: \emph{the equation $g(z)=0$ can not have more
positive roots than $g(z)$ has changes of sign, or more negative roots
than $g(-z)$ has changes of sign}. (It is also convenient to remember
the elementary fact that the product of the roots of a polynomial
equation of odd degree is minus the
constant term.)
\subsection*{\it Case $1$: $B< \frac{1}{3} A(3+2A)$}
In this case $(3AB-3A^{2}-2 A^{3})$ is negative and hence 
$g(z)$ has three changes of sign and $g(-z)$ has
none. Therefore,
there will be either $(1)$ three positive roots of $g(z)$ or $(2)$ two complex
and one positive roots, depending on which condition of (\ref{realcon}) $(A,B)$
satisfies. One of the positive roots in the former case and the
positive root in the latter case will correspond to real positive-definite regular
solutions in the interior.
\subsection*{\it Case $2$: $B> \frac{1}{3}
A(3+2A)$ }
In this case, $(3AB-3A^{2}-2 A^{3})$ is positive and hence $g(z)$ has two changes of sign and $g(-z)$
has one. So $g(z)$ has either $(1)$ two positive roots and one
negative or $(2)$ two
complex and one negative, depending on which condition of (\ref{realcon}) $(A,B)$ 
satisfies. 
\subsection*{\it Case $3$: $B= \frac{1}{3} A(3+2A)$ }
For this special case
$(3AB-3A^{2}-2 A^{3})=0$ and hence the valid
roots are the solutions of  
\begin{equation}
z^{2}-(4A + 3) z + (5 A^{2} +6 A) =0
\end{equation}
which will have either two positive or two complex roots. (The
other root is $z=0$ which is not physically meaningful.) However, neither
of the positive roots, which exist if $A\le \frac{3}{2}$, will be
less than $A$ and hence there is no real infilling solution in the
interior. \\

We now combine Descartes'
rule with the 
conditions of (\ref{realcon}) (Fig. 1). We plot the 
\begin{figure}[!h]
  \begin{center}
    \leavevmode
    \vbox {
      \includegraphics[width=14cm,height=8cm]{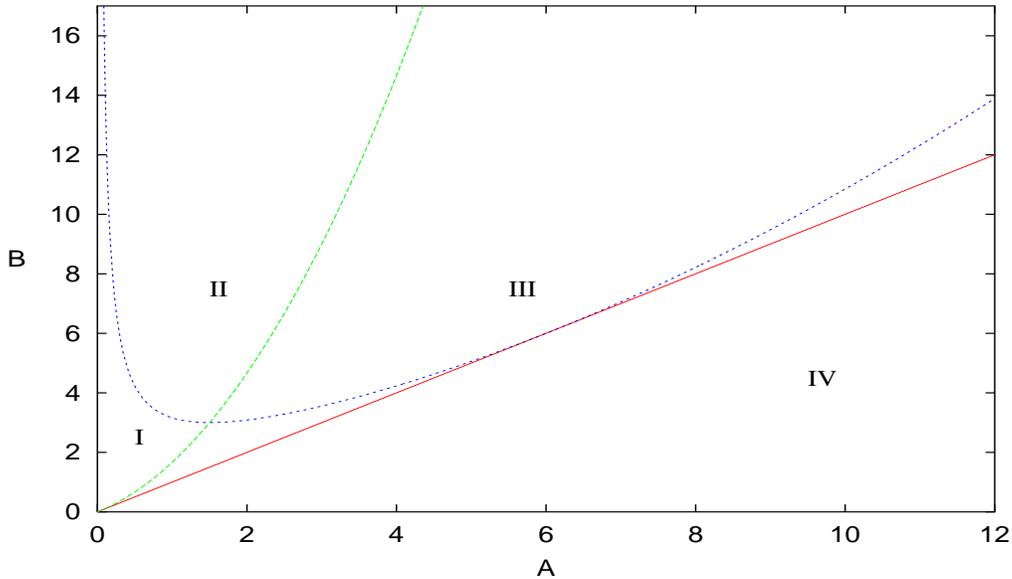}
      \vskip 3mm
      }
    \caption{Structure of Roots (for $\Lambda<0$)} 
    \label{Fig1}
  \end{center}
\end{figure}
curves $B= \frac{1}{3} A(3+2A)$ (green line) and
$B=\frac {4}{81}\,\frac{(A+3)^{3}}{A}$ (blue
line). For
physical interest, we have also plotted the line $A=B$ which represents
isotropy. The two curves divides the $(B,A)$ plane in four regions.
\subsection*{Region I}
This is the region bounded between the two curves on the left-hand side
of their intersection point, with the green line included. One root is negative and two others
positive. The positive roots become equal on the blue line and the
negative root become zero on the green line. None of the positive roots corresponds
to a real infilling
solution in the interior of the Berger sphere.
\subsection*{Region II}
This is the region above the
intersection point of the two curves. So
there is one pair of complex roots and a
negative root. On the green line there are only a pair of complex
roots, the real one being zero. 
\subsection*{ Region III}
Two roots are complex and one is real and positive. The positive root
corresponds
to a real positive-definite infilling
solution in the interior. As above,
on the green line, the real root is zero.
\subsection*{ Region IV}
This is the region bounded by the two lines below their point of
intersection, with the blue line included. All roots are positive and
one of them corresponds
to a real infilling
solution. Two roots are equal on the blue line.
\subsection{Numerical Study}
In the previous section we have studied the general behaviour of the
roots systematically. 
Although $B$ appears only in the constant term of $g(z)$, its role is rather crucial
in determining the sign and reality of the roots. In this section we
will study the roots of $g(z)$ and their corresponding actions numerically, and demonstrate an interesting
connection with catastrophe theory.

First we study the roots and actions for fixed values of
$A$, as functions of $B$. From the analysis in the previous section (and from
Fig. 1), one would expect to study
three distinct generic cases, determined by the value of $A$
less than, equal to and greater than its value where the two curves meet,
i.e., at 
$A=3/2$ (and $B=3$). 
\subsection*{Case 1: $A<3/2$}
For a fixed $A<3/2$, by varying $B$ one moves continuously from region IV (three
positive roots) to the green line where one of the roots become zero (two
still being positive), then into region I, where two roots continue to
be positive, but the other is now negative, and then to the blue line where the
positive roots become equal, and then to region II where they
turn complex (the other still being negative).

The behaviour of the three roots and  the corresponding
Euclidean actions are plotted in Figs. 2 and 3 (as long as they remain
real), as functions of $B$ for $A=0.2$ and $A=1$ respectively. (The same colour is used to show the
correspondence.) 
\begin{figure}[!h]
{\includegraphics[width=0.45\textwidth]{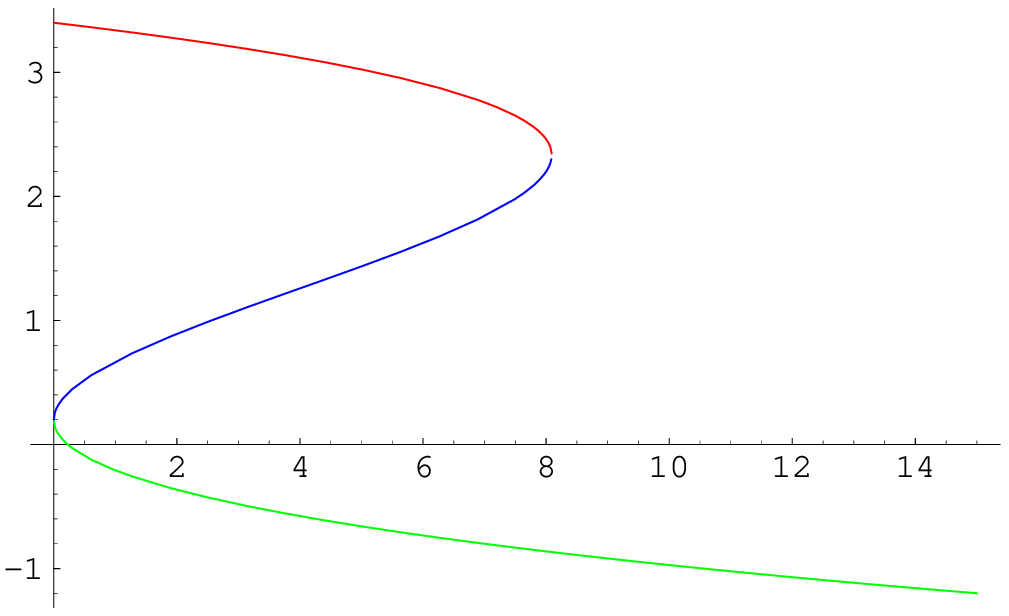}}
\hspace*{1.2cm} 
{\includegraphics[width=0.45\textwidth]{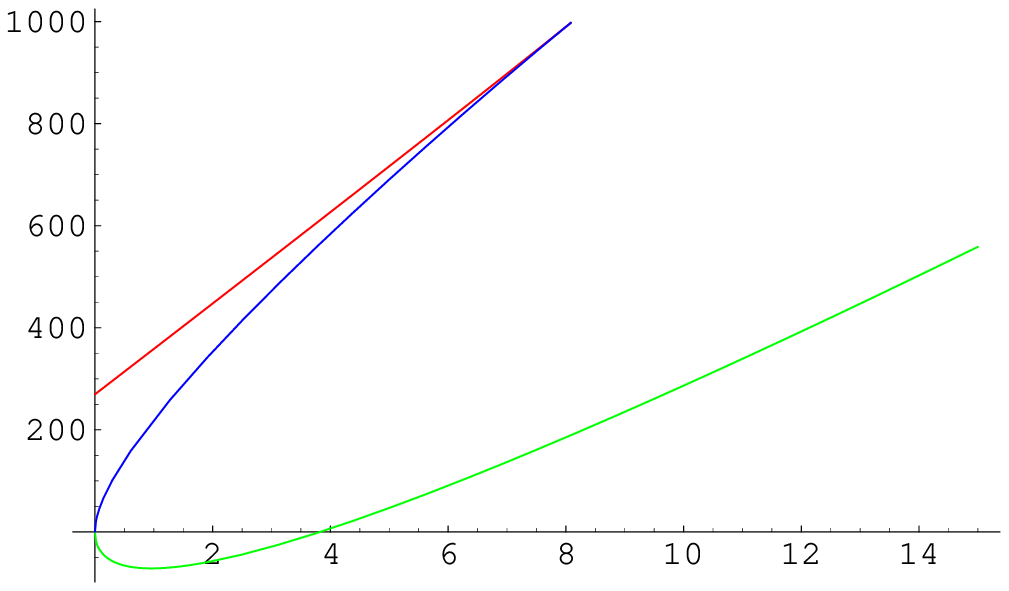}}
\caption{Real Roots and corresponding Actions as functions of $B$ ($A=0.2$)}
\end{figure}
\begin{figure}[!h]
{\includegraphics[width=0.45\textwidth]{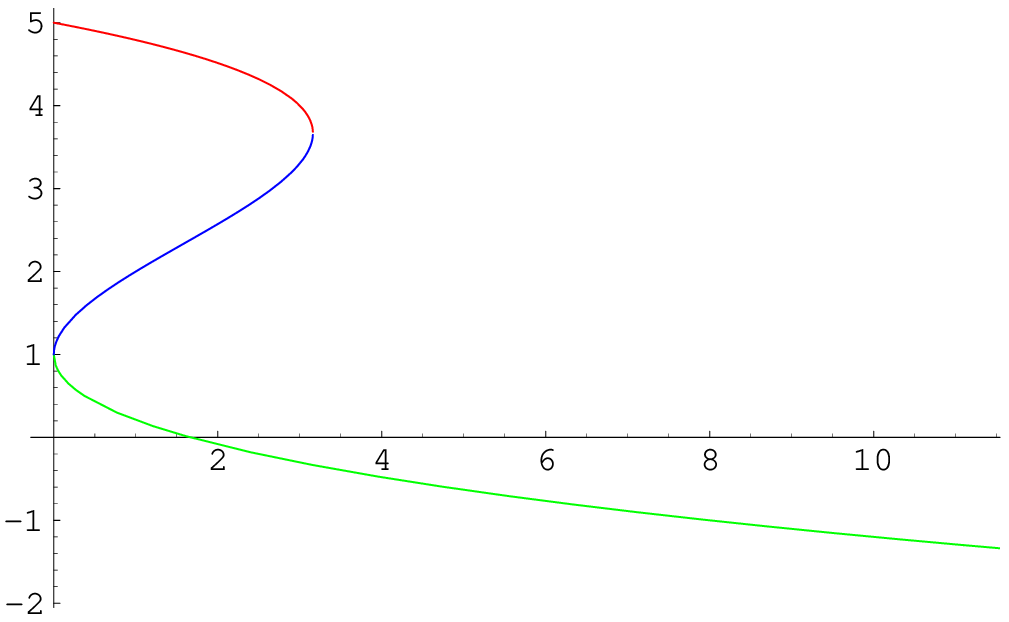}}
\hspace*{1.2cm} 
{\includegraphics[width=0.45\textwidth]{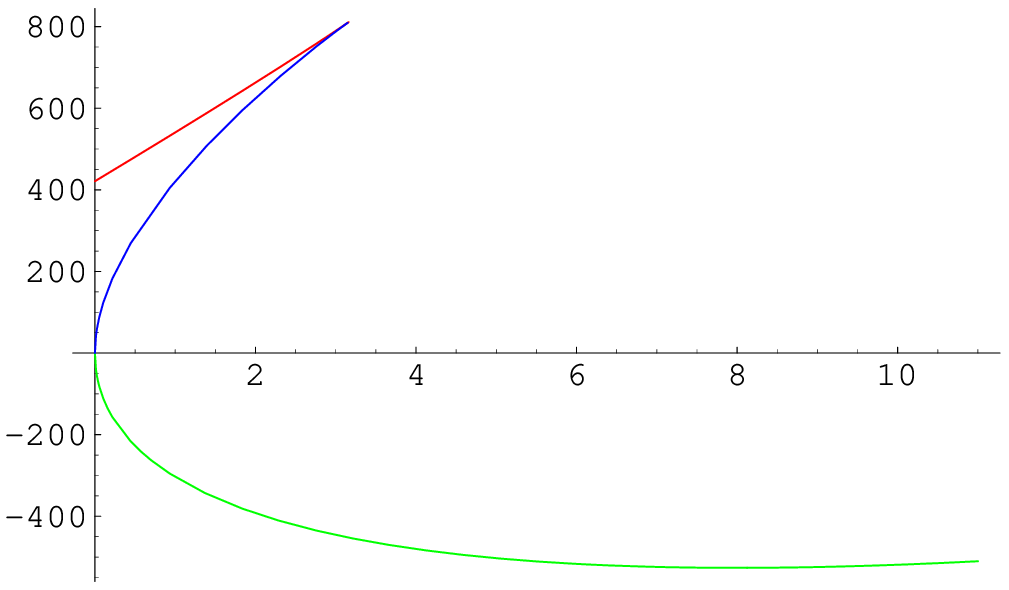}}
\caption{Real Roots and corresponding Actions as functions of $B$ ($A=1$)}
\end{figure}
\subsection*{Case 2: $A=3/2$}
For the special value of $A=3/2$, all roots are positive in
region IV, where, on the point that the green and blue lines meet two become
equal and the other becomes zero. On entering region II the two equal
roots become a complex-conjugate pair and the other turns negative.
The behaviour of the three roots is therefore
similar to the previous behaviour, except that they all turn complex and become
negative at the same value ($B=3$) (Fig. $4$). The corresponding actions
also have a structure similar to those in $A<3/2$.
 \begin{figure}[!h]
{\includegraphics[width=0.45\textwidth]{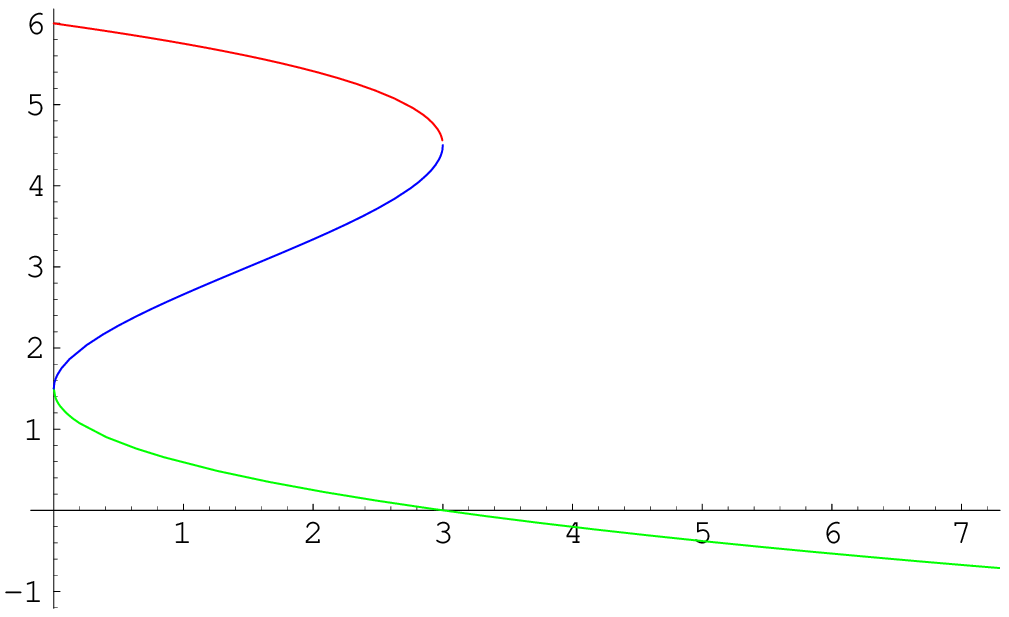}}
\hspace*{1.2cm} 
{\includegraphics[width=0.45\textwidth]{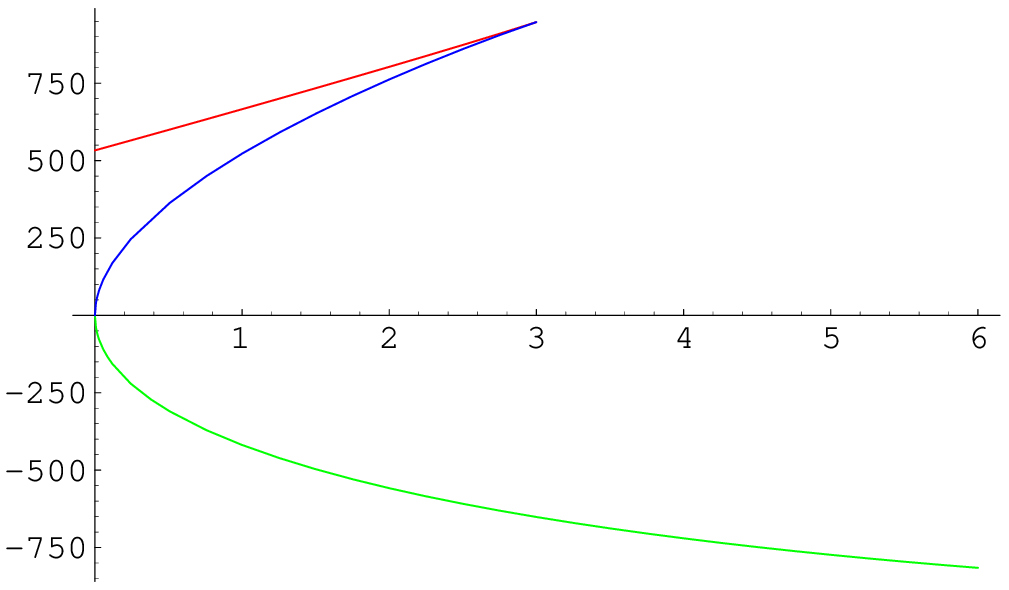}}
\caption{Real Roots and corresponding Actions as functions of $B$ ($A=3/2$)}
\end{figure}
\subsection*{ Case 3: $A>3/2$}
For $A>3/2$ one starts in region IV with three positive roots, two of
which become equal on the blue line and then turn complex and remain so
in region II. The other root remains positive in region III and becomes zero
on the green line to become negative in region II. See Fig. $5$
($A=3$) and Fig. $6$ ($A=100$). 

\begin{figure}[!h]
{\includegraphics[width=0.45\textwidth]{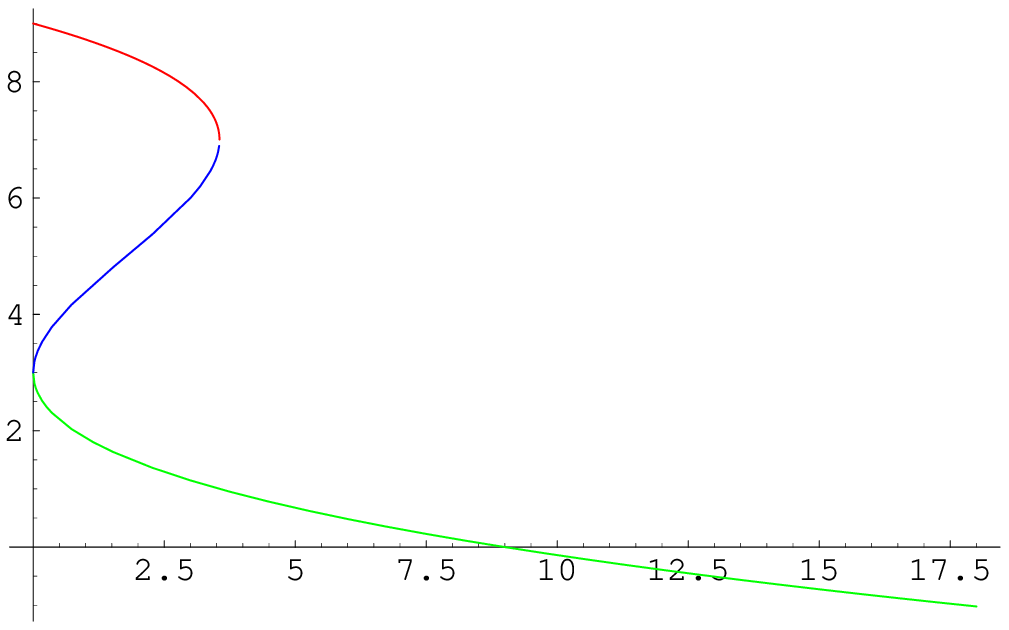}}
\hspace*{1.2cm} 
{\includegraphics[width=0.45\textwidth]{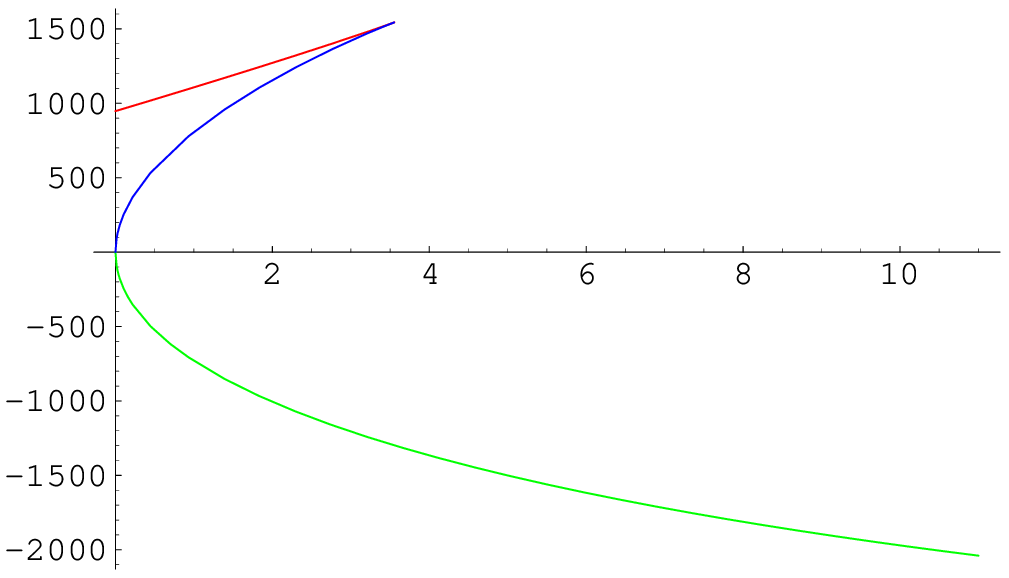}}
\caption{Real Roots and corresponding Actions as functions of $B$ when
$A>3/2$ ($A=3$)}
\label{...}
\end{figure}
\begin{figure}[!h]
{\includegraphics[width=0.45\textwidth]{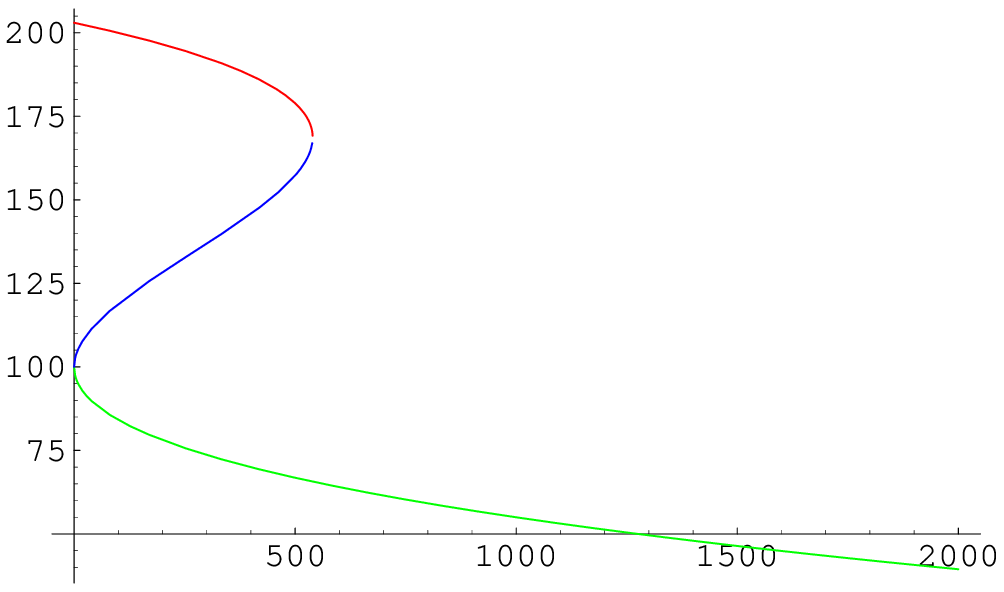}}
\hspace*{1.2cm} 
{\includegraphics[width=0.45\textwidth]{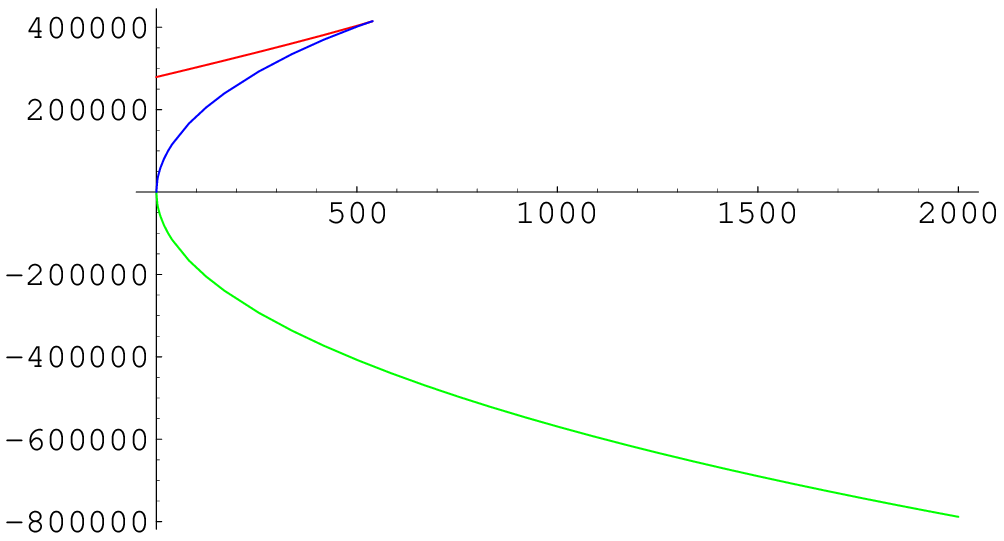}}
\caption{Real Roots and corresponding actions as functions of $B$ for
large $A$($A=100$)}
\label{...}
\end{figure}
\subsection*{The ``Catastrophe Manifold'' of $I_{E}$}
One can see that certain patterns emerge for both $g(z)$ and
$I_{E}$.
For fixed values of $A$, the real solutions of $g(z)$ form
a two-fold pattern as functions of $B$. One can check that this
occurs for all $A$, large and small. The upper and lower folds turn over at 
$B=\frac{4}{81} \frac{(A+3)^3}{A}$ and at $B=0$ respectively. This is
easily understood with the help of Fig. $1$: the curves (of two roots) in
the upper fold meet when the two roots become equal at $B=\frac{4}{81}
\frac{(A+3)^3}{A}$, i.e., before entering the combined region of II and III where they turn
complex. On the other hand, on $B=0$, $g(z)$ has a double root equal to $A$ (the other root
is $2A$) -- therefore the lower folds turn over around the $B=0$
line. The surface $g(z)=0$
thus formed by placing such
images successively (Fig. $11$) is similar in structure to a cusp catastrophe manifold
familiar in dynamical systems driven by a quartic potential with two
control parameters (see, for example, \cite {PS}).
The minima of a quartic potential occur
when its derivative (a polynomial of degree three) is set to zero and hence the catastrophe
manifold represents the equilibrium points of the system. The ``catastrophe
map'' is then the part of projection of the catastrophe manifold
onto the plane of the control variables bounded between
the lines along which the folds turn over -- in our case this is the
region bounded between $B=0$ and $B=\frac{4}{81}
\frac{(A+3)^3}{A}$ curves in Fig. $1$, i.e., the union of region I and
IV. However, one may
wonder why a
cusp does not appear in this case. This is because of the (particular)
way in which
$A$ and $B$ combine in the coefficients of $g(z)$ and also because they are
constrained to be positive -- both facts are dictated by the
physical configuration. One can 
obtain a cusp, however, by working with the new variable
$D(=AB)$ instead of $B$. A cusp will appear at $A=-3$, which is obviously outside the
range of our physical interest. 

It is not difficult to see that $I_{E}$ has the same catastrophe map,
namely the union of region I and
IV. However, the more important observation is that the
relative orientation of the folds of $I_{E}$ is the same as that of
the corresponding roots of $g(z)$, i.e., they do not cross each other. This pattern persists even when one gets very close
to $A=0$ as shown in Fig. $7$. For very small values of $A$ the green and blue lines nearly overlap and
coincide completely for $A=0$. They therefore meet the red line at
infinity. For
higher values of $A$ one gets a persistent behaviour as in Figs. $8$-$10$. Therefore the
``catastrophe
manifold'' of
$I_{E}$ is ``diffeomorphic'' to the one found above for $z$, i.e., the
surface $I_{E}$ is obtainable from that of $z$ by
a deformation which preserves the catastrophe map. This is not
automatic or obvious given the form of $I_{E}$ which is a quadratic
function of $z$. Note, however, that the surface of $I_{E}$ is not
smooth where the upper fold occurs.

The two-dimensional
catastrophe map in a dynamical system with a quartic potential
demarcates the regions of
three stable minima and one stable minimum. In our
case they indicate a demarcation between the regions where there are
three real $I_{E}$ and one real $I_{E}$ for the boundary-value
problem. 
\begin{figure}[!h]
{\includegraphics[width=0.28\textwidth]{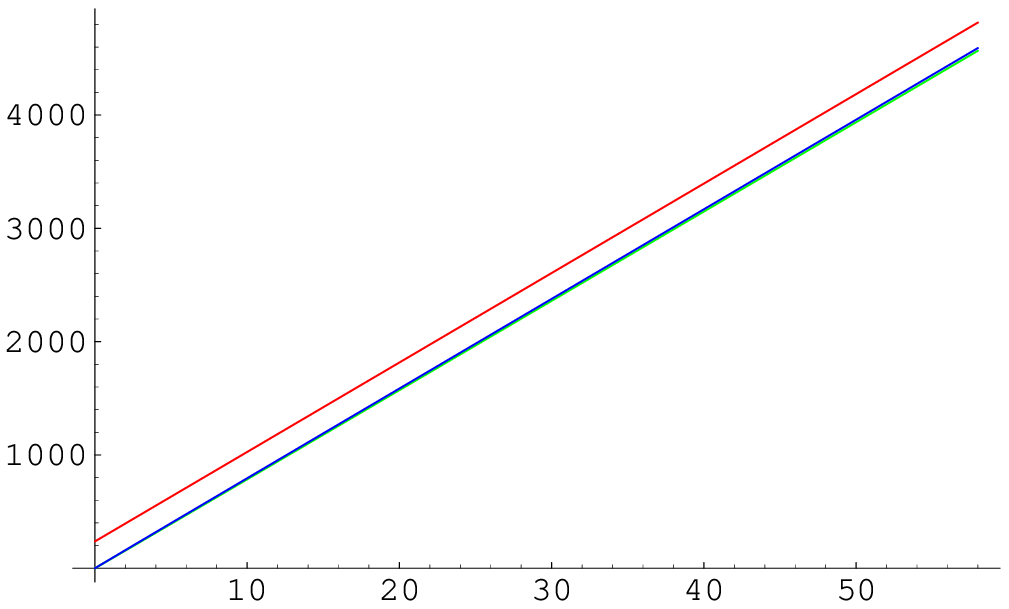}}
\hspace*{1.0cm} 
{\includegraphics[width=0.28\textwidth]{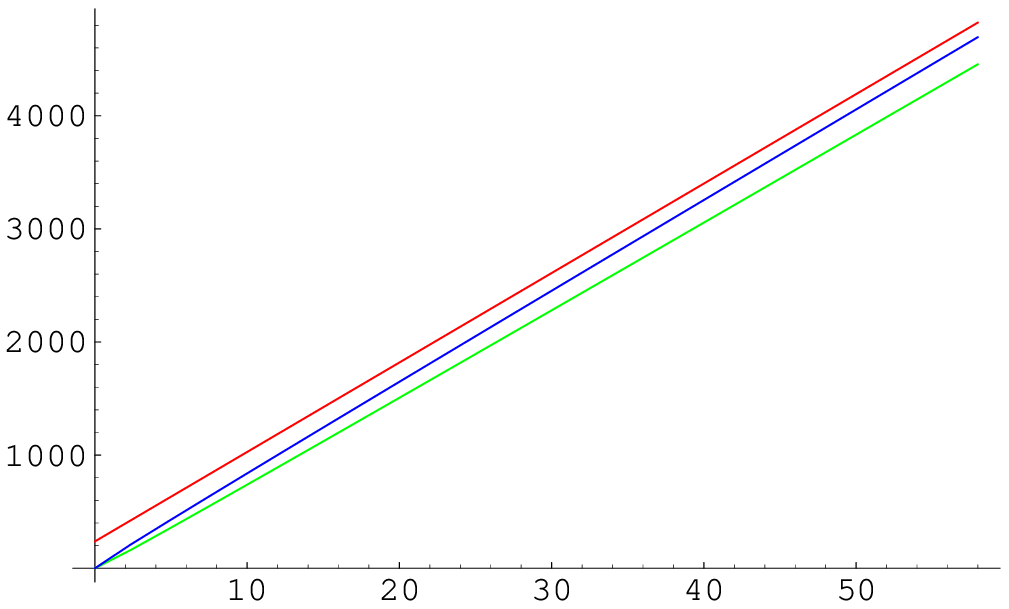}}
\hspace*{1.0cm} 
{\includegraphics[width=0.28\textwidth]{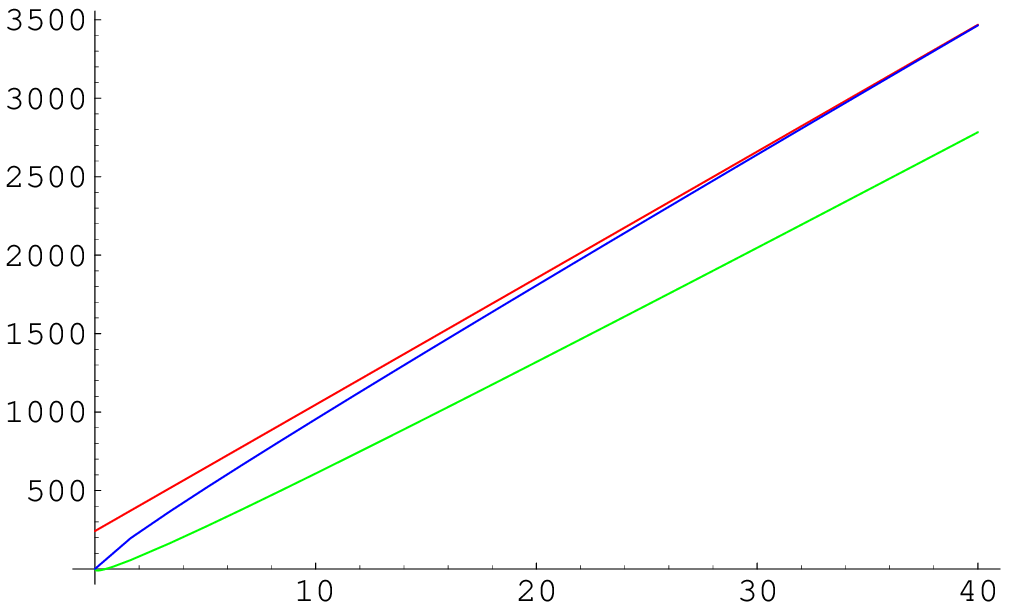}}
\caption{Actions as functions of $B$ ($A=25\times 10^{-6},\,25\times
10^{-4},3\times 10^{-2}$)}
\label{...}
\end{figure}
\begin{figure}[!h]
{\includegraphics[width=0.28\textwidth]{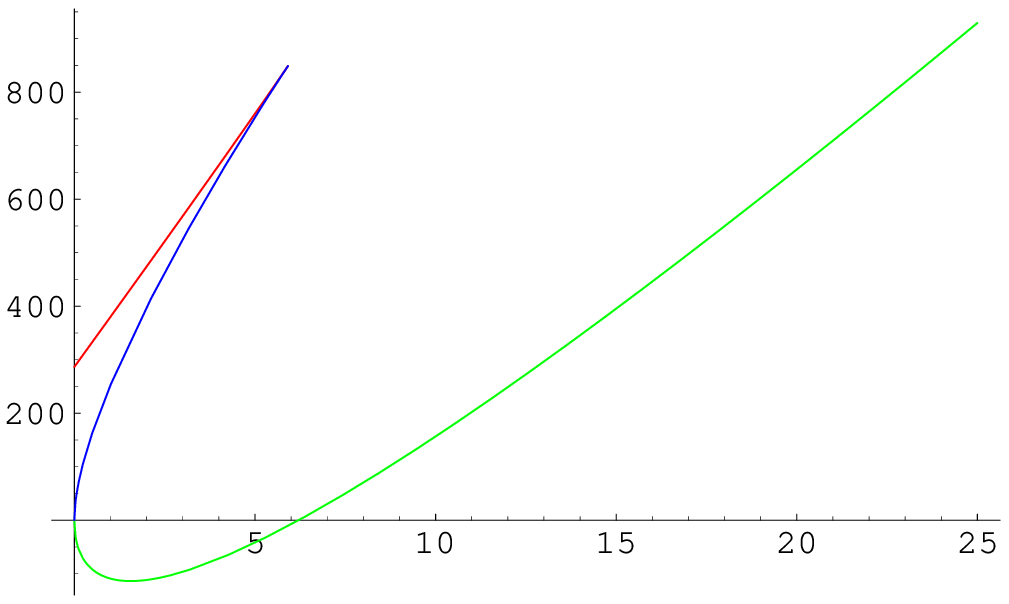}}
\hspace*{1.0cm} 
{\includegraphics[width=0.28\textwidth]{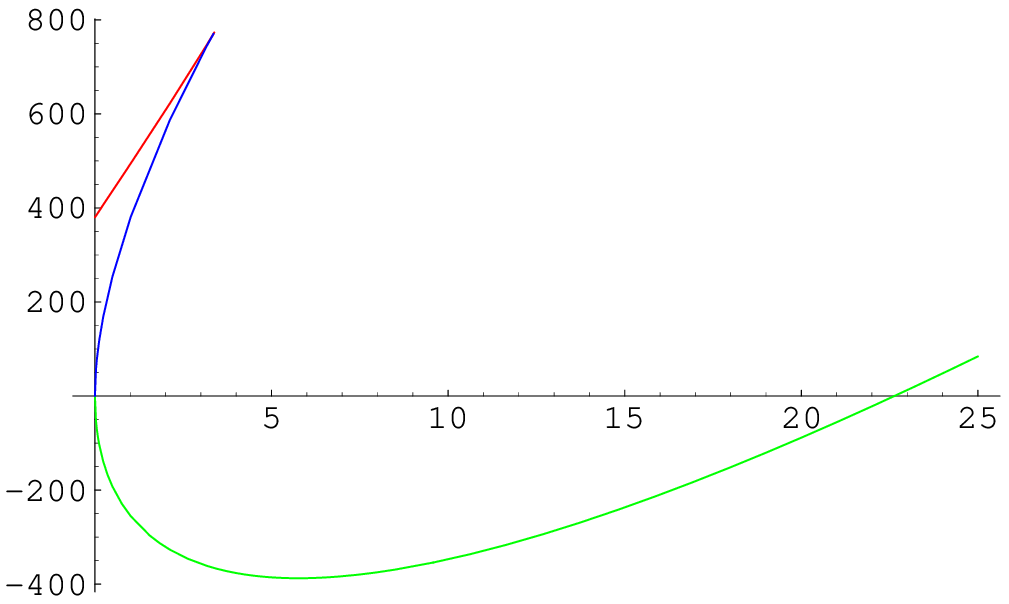}}
\hspace*{1.0cm} 
{\includegraphics[width=0.28\textwidth]{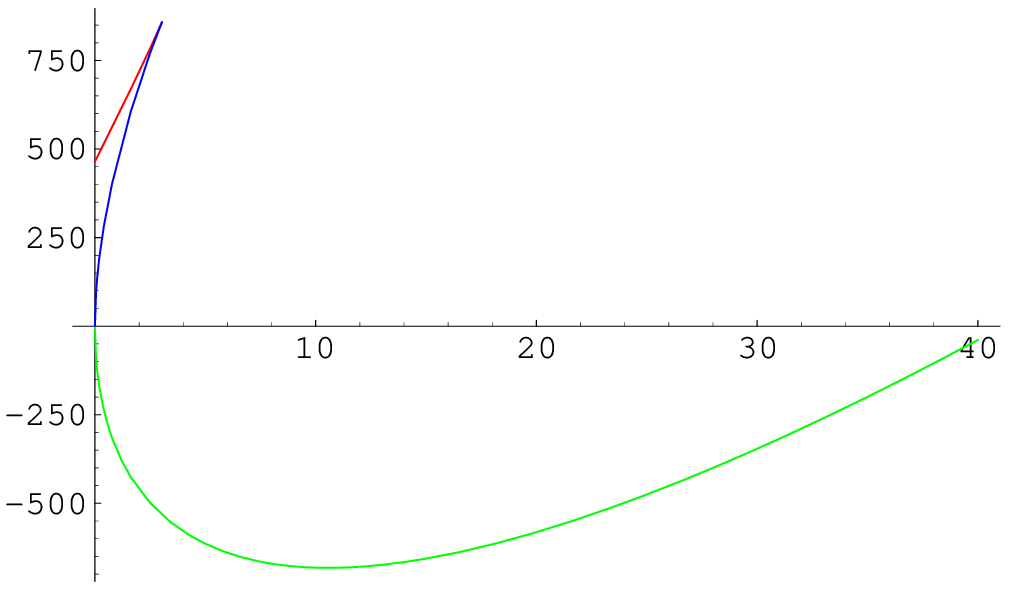}}
\caption{Actions as functions of $B$ ($A=3\times 10^{-1},\,8\times 10^{-1},\,1.2$)}
\label{...}
\end{figure}
\begin{figure}[!h]
{\includegraphics[width=0.28\textwidth]{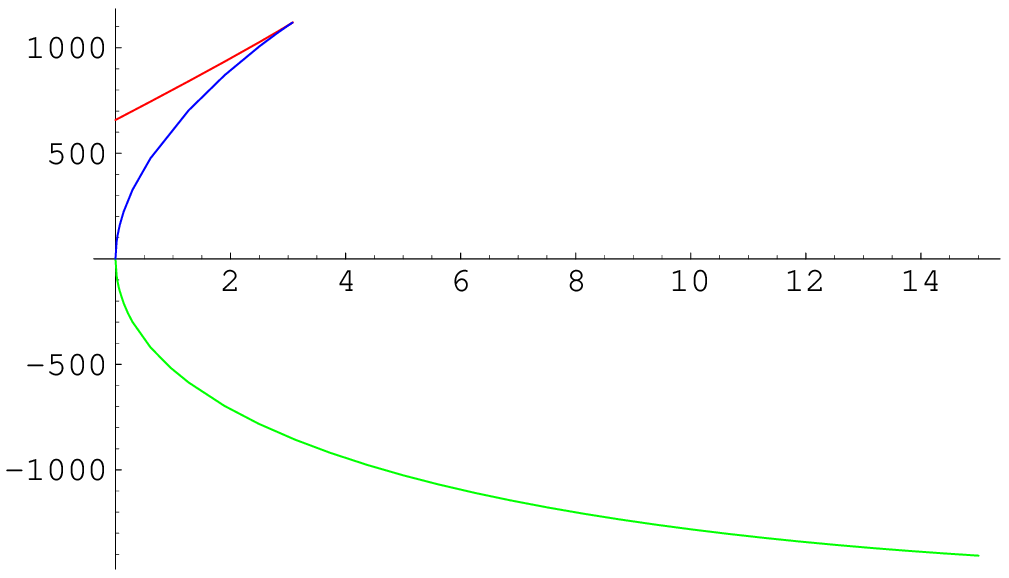}}
\hspace*{1.0cm} 
{\includegraphics[width=0.28\textwidth]{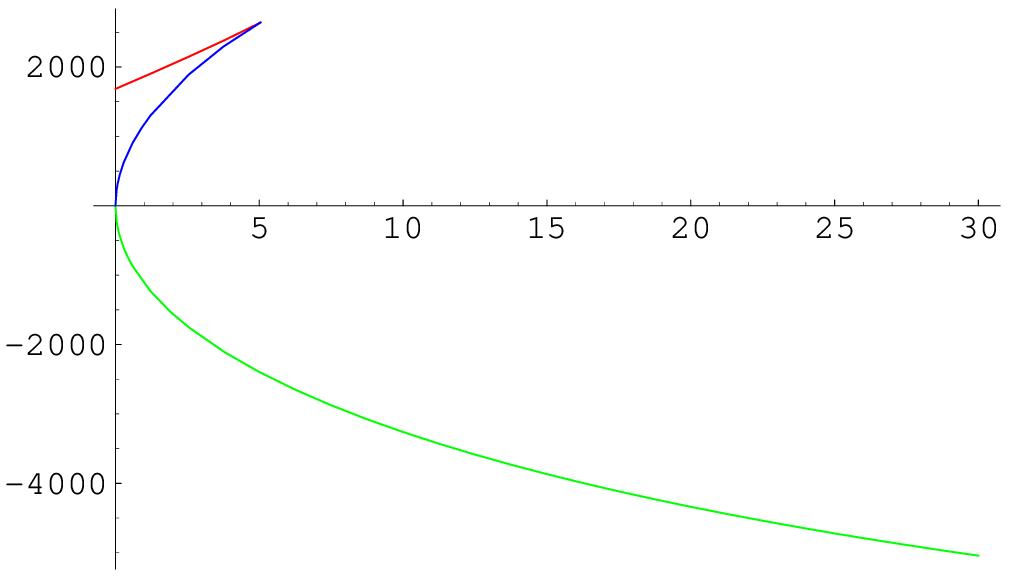}}
\hspace*{1.0cm} 
{\includegraphics[width=0.28\textwidth]{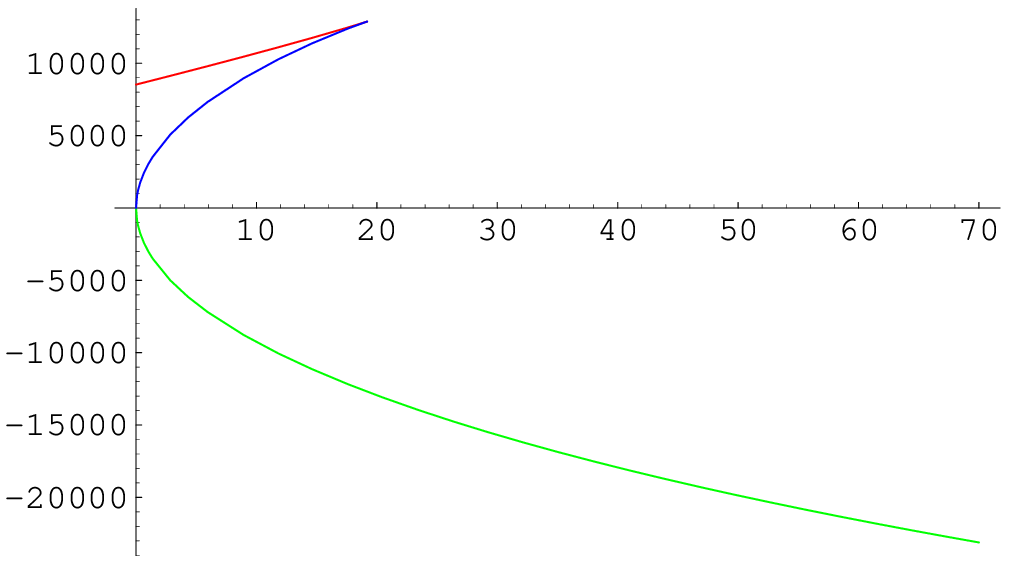}}
\caption{Actions as functions of $B$ ($A=2.0,\,5.0,\,15.0$)}
\label{...}
\end{figure}
\begin{figure}[!h]
{\includegraphics[width=0.28\textwidth]{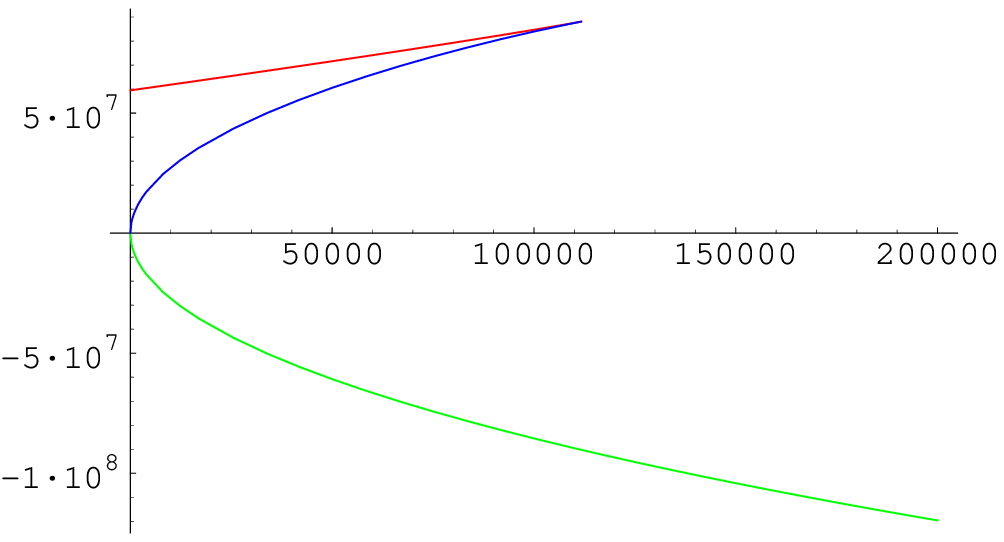}}
\hspace*{1.1cm} 
{\includegraphics[width=0.28\textwidth]{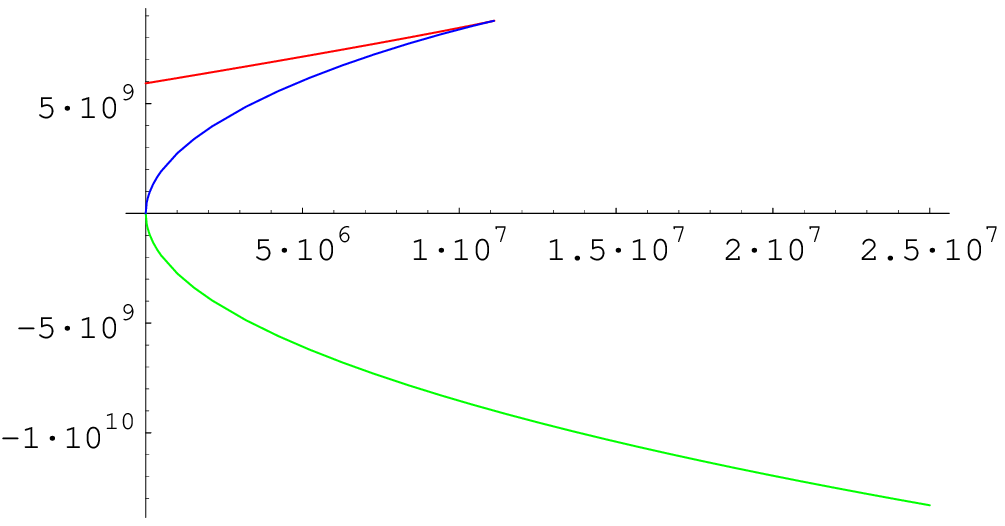}}
\hspace*{1.1cm} 
{\includegraphics[width=0.28\textwidth]{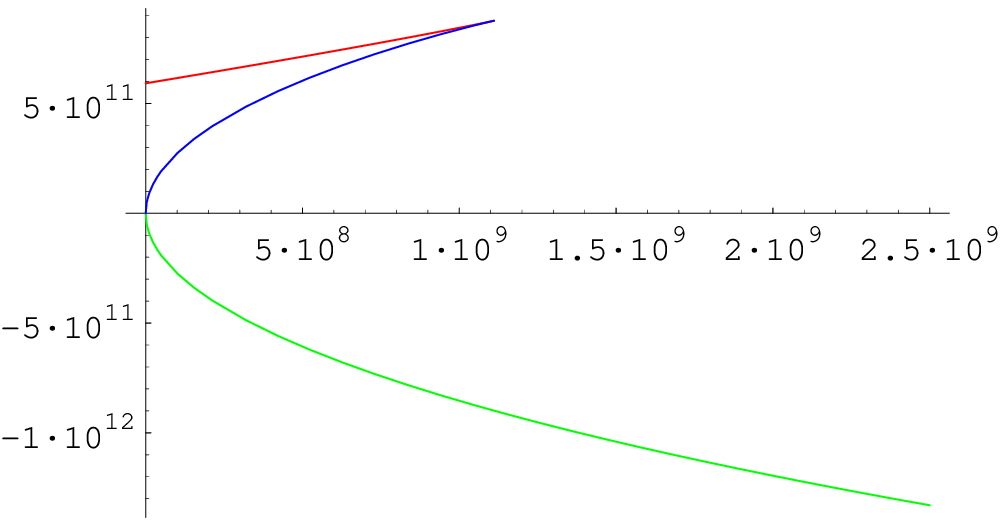}}
\caption{Actions as functions of $B$ ($A=1500,\,15000,\,150000$)}
\label{...}
\end{figure}
\begin{figure}[!h]
  \begin{center}
    \leavevmode
    \vbox {
      \includegraphics[width=9cm,height=8.2cm]{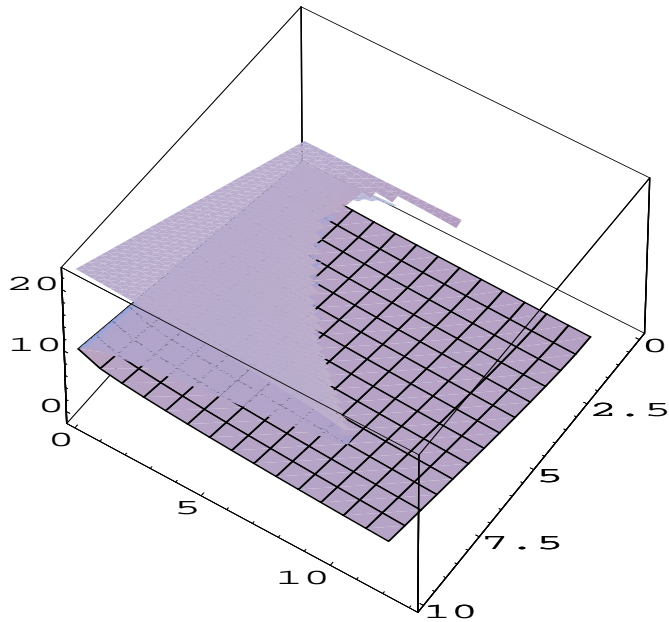}
      \vskip 3mm
      }
\caption{The ``catastrophe manifold'' $g(z)=0$ : the
    surface of $I_{E}$ is similar in shape and
    orientation except at the upper fold where it is not smooth.} 
      \end{center}
\end{figure}
\subsection{Large Radii and Small Anisotropy}
The observations that the space of $I_{E}$ is diffeomorphic to that
of $z$ and that it does not intersect itself have immediate physical
implications. This means that the 
dominant contribution will always come from {\it{one}} solution, namely the
one represented by the green curve in Figs. $2$-$10$ (and the meshed
surface in Fig. $11$). Note that this is the
solution which gives a positive-definite
infilling solution as long as 
$b^2< \frac{1}{3}a^2\,(2a^2+3)$. One can verify from Figs.
$2$-$6$ that this solution always takes values within the interval $(0,A)$.

In cosmology, one is more interested in regions where $b$ is not
greatly different from $a$ and usually when they are both
large. All three roots are positive in this latter
region. Taking $A=B$, the roots of
$g(z)$ are 
\begin{equation}
\begin{array}{rcl}
z_{1}&=&A+\frac{3}{2}-\frac{1}{2}\,\sqrt {12\,A+9},\\
z_{2}&=&2\,A,\\
z_{3}&=&A+\frac{3}{2}+\frac{1}{2}\,\sqrt {12\,A+9}.\\
\end{array}
\end{equation}
It is easy to check that only $z_{1} < A$ and, hence, $z_{1}$ corresponds to the positive-definite real infilling
solution in the interior. The dominant contribution to the path
integral (\ref{1}) will therefore come
from $z_{1}$ (which can also be checked explicitly in this case). The
action of this solution is given by 
\begin{equation}
8 \pi \,G \, (\lambda I_{E})= -\frac{4}{3}\,{\pi}^{2}\left (-9+3\,\sqrt {12\,A+9}+4\,A\sqrt {12\,A+9}\right 
)\,\, \sim\,\, -\frac{32}{{\sqrt{3}}}\,
{\pi}^{2}\,A^{\frac{3}{2}},
\end{equation} 
so becoming more negative as $A$ grows. Actions for $z_{2}$ and
$z_{3}$ are positive and become more positive as $A$ grows, as can be
checked explicitly by direct substitution.
\section{Conclusion}
We have shown that, for a given boundary which is a Berger sphere
(a squashed $S^{3}$ with two axes equal), there are in general
three distinct ways in which one can fill in with a self-dual 
Taub-NUT-(anti)de Sitter metric.
With
suitable choice of variables, the problem of finding explicit
solutions for such infilling metrics
can be translated into a univariate algebraic equation of degree three
and hence can be solved exactly. 
The Euclidean action $I_{E}$ is a quadratic function of the
solutions of this third-degree equation and hence, corresponding to
every solution of this equation,
the action can be
found purely in terms of the boundary data -- the two radii
$(a,b)$ of the Berger sphere, for both positive and negative
cosmological constants. The positive and negative roots of this
equation correspond to metrics for which actions are real -- only
complex solutions correspond to complex-valued actions in
general. 

In the case of a negative cosmological
constant, we have further
discussed systematically the ranges of $a$ and $b$ where the
solutions of the algebraic system lead to real- and complex-valued
solutions in the interior and have studied the structure of the three
roots. We found that one of the roots corresponds to a
positive-definite infilling solution if and only if $b^2<
\frac{1}{3}a^2\,(2a^2+3)$. 
For small squashing, i.e., when $a$ and $b$ are of the same order
and not too small, all three roots are positive. When $a$ and $b$ are exactly equal, this holds
for small radius as well. We therefore investigated the
roots and their corresponding
actions 
numerically in this range (i.e., until two of them turn complex) as
functions of the boundary variables $a$, $b$.
 We found that both the roots
and the actions have structures
similar to those of the cusp catastrophe in dynamical systems. The
``catastrophe manifold'' of $I_{E}$ does not intersect itself
which 
implies that the the dominant contribution will come from
the positive-definite infilling solution.
Further,
the classical actions for large values of the radii $(a,b)$ in their
isotropic limit (of particular interest for cosmology) have been
discussed. The (dominant) contribution coming from the positive-definite
solution has an
action proportional to $-a^3$, thereby favouring large radii.
\subsection*{Acknowledgements}
We would like to thank Gary Gibbons, Stefano Kovacs, Alexei Kovalev,
Jorma Louko, Henrik Pedersen, Nick
Shepherd-Barron and Galliano Valent for helpful discussions and comments. MMA was supported
by awards from the Cambridge Commonwealth Trust and the Overseas Research Scheme.

\end{document}